\documentclass[amssymb,twocolumn,aps]{revtex4}
\usepackage{graphics,color,graphicx,amsmath}
\usepackage{subcaption}
\usepackage{natbib}
\usepackage{verbatim}
\usepackage{graphicx}
\usepackage[above,below]{placeins}
\usepackage{float}
\usepackage{tabularx}
\usepackage{times}
\usepackage{blindtext}
\usepackage{url}
\usepackage{rotating}
\usepackage{mathtools}
\usepackage{threeparttable}
\usepackage{enumitem}
\setlist{nosep}   

\renewenvironment{quote}
  {\list{}{\rightmargin=.1cm \leftmargin=.3cm}%
   \item\relax}
  {\endlist}

\begin{document}

\title{``Existential dread": Student and faculty views on the impact of changes in federal science funding to physics research
}

\author{Meagan Sundstrom~\footnote[1]{sundstrommeagan@gmail.com}, Ian Olivant, and Eric Brewe~\footnote[2]{Corresponding author: eb573@drexel.edu}}
\affiliation{Department of Physics, Drexel University, Philadelphia, Pennsylvania 19104, USA}

\date{\today}

\begin{abstract}
Sustained investment in physics research is essential not only for advancing science and technology, but also for training the next generation of scientists, including undergraduate students, graduate students, and postdoctoral researchers. Through their research experiences with faculty advisors, these trainees develop critical skills such as experimental design, data analysis, scientific communication, and collaborative problem-solving. Starting in early 2025, the United States administration introduced changes that affected the funding landscape for physics research: grant funding was slashed, grant review panels were delayed or canceled, and some universities froze hiring and research spending. This study aims to better understand the impacts of these changes to academic physics researchers at different career stages. Using survey responses from physics undergraduate students, graduate students, postdoctoral researchers, and associate professors, we find that most academic researchers perceive the impacts to their research as negative. Some of the most commonly reported impacts include loss of existing research funding, concern about obtaining research funding in the future, and job insecurity. We discuss differences in these impacts across career stages and potential implications for the research training pipeline in physics. 
\end{abstract}

\maketitle

\section{Introduction}

Starting in early 2025, the United States administration implemented substantial changes to the federal science funding landscape. In response to executive orders, federal agencies such as the National Science Foundation (NSF) paused new awards, delayed reviews for pending proposals, and vetted existing grants for compliance~\cite{impey2025funding,crockett2025nihnsf,mervis2025nsfcrisis,gewin2025stemdivefforts,mervis2025nsfvetting}. Although Congress passed appropriations in early 2026 that largely restored funding levels, agencies continue to face delays in disbursing budgets~\cite{mervis2026fundingflow} and are continuously undergoing personnel reorganizations~\cite{mervis2025nsfreorg}. Policies restricting employment opportunities for international scholars also remain in effect~\cite{mervis2025visa}.

The changing funding landscape has myriad consequences for researchers at colleges and universities. Many early-career scientists are relocating to other countries~\cite{witze2025scientistsleaving,aksenfeld2025earlycareer,gibney2026france} as they face uncertainty about their career trajectories~\cite{langin2025earlycareer,langin2026pressure,AzoulaySadunScur2026}. Graduate program admissions and enrollment have decreased, particularly for international students~\cite{Mulvey2026}. Opportunities for undergraduate students to develop scientific skills through mentored research have been halted~\cite{voosen2025nasaH2O,mervis2025nsfREU,WuEtAl2026}. Disruptions at the federal level, therefore, have ripple effects across the entire academic pipeline and directly affect undergraduate and graduate training and postdoctoral opportunities.

Such disruptions may affect physics research in distinct ways from research in other scientific disciplines. One reason is that each discipline relies on different funding sources. Physics researchers often receive support from the NSF, the Department of Energy, the Department of Defense, and the National Aeronautics and Space Administration. Other scientific researchers, for example in chemistry and biology, are also funded by the National Institutes of Health and biotech industry partnerships. Delays and cuts in federal appropriations may affect physics differently than fields supported by other funding ecosystems. The alignment between research topics and current administrative priorities also differs across fields, shaping the particular ways in which researchers may be affected by the changing funding landscape. Biologists, for example, have faced targeted cuts to climate-related research~\cite{Lackner2016PromiseNegativeEmissions} and research using sex as a biological variable~\cite{Becker2025SexDifferences}. Meanwhile, quantum information science and artificial intelligence, two central research areas for physicists, remain national strategic interests that are less at risk for funding cuts~\cite{mckenzie2025ai}. It is important to understand the effects to researchers in physics specifically in order to support the development of future physicists and uphold the national physics research enterprise.

Several existing reports have used rapid response or informal surveys to measure the impacts of recent funding changes to physics students and researchers in higher education settings~\cite{white2025impacts,bennett2025advocacy,Hammond2025trump}. The American Institute of Physics, for example, surveyed department heads about the future of physics and astronomy graduate programs~\cite{white2025impacts}. About a third of the 115 respondents anticipated fewer incoming PhD students in the upcoming years. Roughly half of the respondents reported that faculty in their department are experiencing or anticipating cuts in their federal funding, with these effects disproportionately impacting private institutions more than public institutions. On open-ended survey questions, department heads reported a range of effects due to the changing funding landscape, including ``demoralization of students and faculty, negative effects of hiring freezes brought about by funding uncertainty, and anxiety for students and faculty resulting in decreased well-being and productivity. Some also expressed considerable fear of the effects of changes in immigration and visa policy creating a brain drain as students and faculty look for opportunities in other countries" ~\cite{white2025impacts}.

Another research team surveyed American Physical Society (APS) members about the extent to which APS has supported their needs since the 2025 inauguration~\cite{bennett2025advocacy}. The survey included open-ended questions where respondents could elaborate on the impacts of federal actions to their professional lives. The 100 APS members who responded shared several common concerns, including ``a general hopelessness and anxiety," fear for students in the United States on visas, and ``lower interest in attending conferences to which they could not drive"~\cite[p. 7]{bennett2025advocacy}. 


The above investigations establish a baseline understanding of the impacts of recent federal changes in science funding to physics researchers at colleges and universities. In the current study, we build on this understanding in several ways. First, we purposefully sample from students and researchers across a range of career stages (i.e., undergraduate students through associate professors) rather than only one. This sampling ensures that our findings capture a comprehensive picture of how federal funding changes affect the entire academic research ecosystem. Second, we collect survey data systematically rather than through rapid response sampling. That is, we design our data collection procedures to achieve high response rates within each population (i.e., career stage) to ensure that the results are representative. The research questions driving our exploratory study are: 
\begin{enumerate}
    \item How do physics undergraduate students, graduate students, postdoctoral researchers, and faculty members perceive the effects of federal changes in science funding to their research (e.g., positive, negative, or no impact)? 
    \item What specific impacts to research do physics undergraduate students, graduate students, postdoctoral researchers, and faculty members identify as resulting from federal changes in science funding?
\end{enumerate}

\section{Methods}

We collected and analyzed survey data from three different populations of physics researchers: associate professors, graduate students and postdoctoral researchers, and undergraduate students. Below, we describe the survey items we administered, the populations and contexts from which we collected the survey data, and our data analysis procedures.

\subsection{Survey items}

To address our research questions, we used two survey items---a Likert item and an open-ended item---developed in prior work~\cite{AzoulaySadunScur2026}. The Likert item probed researchers' general perceptions of the changes in the funding landscape, whether positive or negative:
\begin{quote}
\textit{There have been major changes in science funding recently as a result of federal executive orders. What has been the impact on your research?
\begin{itemize}
    \item Very negative
   \item Negative
   \item No impact
   \item Positive
   \item Very positive
   \item NA--I am not actively doing research.
\end{itemize}}
\end{quote}
The open-ended item immediately proceeded the Likert item and asked respondents to describe the impacts of the federal changes to their research:
\begin{quote}
\textit{Please elaborate on the impacts on your research, if there have been any. Any additional context you can provide would be very helpful.}
\end{quote}
The prompt was followed by an open text box in which respondents could write as much as they wanted.


Described further below, these survey items were part of longer surveys and were not required: respondents could leave either item or both items blank if they did not understand the prompt or preferred not to answer.

\begin{table*}[t]
\caption{Summary of data sources analyzed in this study. All populations are in the discipline of physics. All respondents to the open-ended item also responded to the Likert item. \label{info}}
\setlength{\tabcolsep}{2
pt}
\renewcommand{\arraystretch}{1.3}
\begin{tabular}{llllll}
\hline
\hline
Population &
Survey context &
Survey timing &
Total survey respondents &
\parbox[t]{2cm}{\raggedright Respondents to Likert item}   &
\parbox[t]{2.5cm}{\raggedright Respondents to open-ended item} \\
\hline
\parbox[t]{3cm}{\raggedright Associate professors} 
& \parbox[t]{4cm}{\raggedright Gordon and Betty Moore Foundation EPI Initiative}
&  July 2025
&  55
&  54
&  44
\\ 
\parbox[t]{3cm}{\raggedright Graduate students and postdocs} 
& \parbox[t]{4cm}{ \raggedright APS Conference for Graduate Women and Gender Minorities in Physics}
&  September 2025
&  101
&  96
&  54
\\
\parbox[t]{3cm}{\raggedright Undergraduate students}  
& \parbox[t]{4cm}{\raggedright APS Conference for Undergraduate Women and Gender Minorities in Physics}
&  January 2026
&  152
&  104
&  49
\\
\hline
\end{tabular}
\end{table*}

\subsection{Data sources}

We administered these two items on surveys collected from three distinct populations who conduct physics research: associate professors of physics, physics graduate students and postdoctoral researchers (hereon referred to as ``postdocs"), and physics undergraduate students (Table~\ref{info}). The authors of this paper are external evaluators for national programs aimed at each of these populations, which facilitated the data collection.

The associate professors were all funded investigators of the Gordon and Betty Moore Foundation's Experimental Physics Investigator (EPI) Initiative. The EPI provides five years of funding, up to \$1.3 million, to associate professors conducting experimental physics research in any subfield. Investigators are tenured faculty at a United States institution who are no more than five years beyond their tenure appointment. The goals of the program include fostering cohorts of distinguished scholars and supporting researchers' creative and novel ideas at a time when their early-career and/or start-up funding may have run out. We included the survey items as part of a longer online survey administered by the Foundation in July 2025. The survey was sent to three cohorts of investigators, for a total of 56 individuals. 55 associate professors responded to the survey, 54 of whom responded to the Likert item and 44 of whom responded to the open-ended item (all respondents to the open-ended item also responded to the Likert item). We did not collect data from assistant or full professors of physics because the EPI initiative is only for associate-level professors and we could not readily access an expansive network of these other populations.

The graduate students and postdocs were all participants of the 2025 Conference for Graduate Women and Gender Minorities in Physics, held in Washington, DC. The two-day conference is organized by the APS and aims to bring together individuals from historically marginalized gender identities in physics. There are sessions dedicated to job applications, mentoring, and networking, among others. Participants pay a \$45 registration fee, with meals, lodging, and travel costs fully covered by the APS. We added the two survey items to a longer, online conference registration survey which  included other questions about participants' experiences, attitudes, and career intentions in physics. The registration survey was administered in September 2025, a few weeks before the conference, and all conference participants were required to fill it out. 101 graduate students and postdocs completed the survey, 96 of whom responded to the Likert item and did not select ``NA--I am not actively doing research" and 54 of whom responded to the open-ended item (all respondents to the open-ended item also responded to the Likert item). We note that 10 of the Likert item responses and 6 of the open-ended item responses from this population were from postdocs. While there are distinguishing features between graduate students and postdocs, especially in the context of scientific funding, we combined the two career stages into one category in our analysis. We wanted to preserve participant anonymity given that the postdoc sample sizes are small and the survey context is identity-based, increasing the possibility of identification.

The undergraduate students were all participants of the 2026 Conference for Undergraduate Women and Gender Minorities in Physics, held in Denver, Colorado immediately preceding the APS Global Physics Summit. Similar to the conference for graduate women and postdocs described above, the two-day conference is organized by the APS and aims to bring together individuals from historically marginalized gender identities in physics. There are sessions dedicated to graduate school applications, navigating male-dominated spaces, and networking, among others. Students have the option to present a research poster as well. Participants pay a \$45 registration fee, with meals, lodging, and travel costs fully covered by the APS. We added the two survey items to a longer, online conference registration survey which included other questions about participants' experiences, attitudes, and career intentions in physics. The registration survey was administered in January 2026, about a month before the conference, and all conference participants were required to fill it out. 152 undergraduate students completed the survey, 104 of whom responded to the Likert item and did not select ``NA--I am not actively doing research" and 49 of whom responded to the open-ended item (all respondents to the open-ended item also responded to the Likert item).

\begin{figure*}[t]
    \centering
    \includegraphics[width=7in, trim = {0 0.65cm 0 0}]{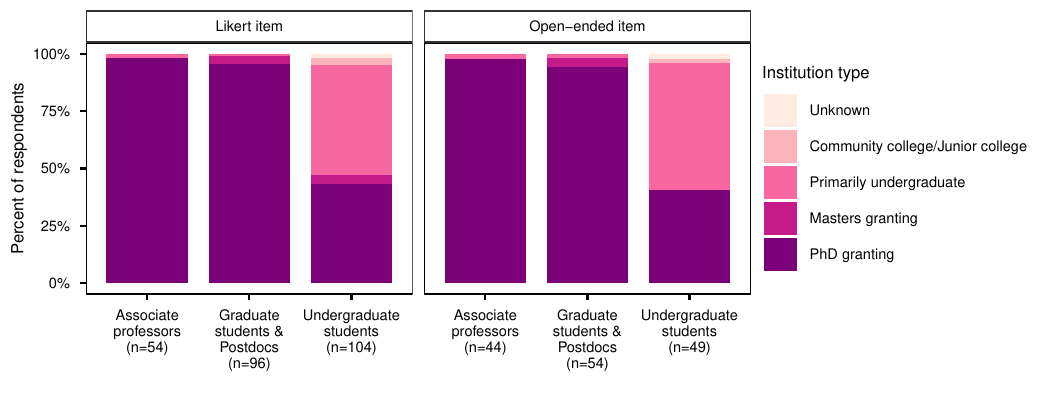}
    \caption{Percent of respondents to each survey item who are affiliated with each institution type.
    }
   
    \label{fig:insttypes}
\end{figure*}

Given these national survey contexts, the data come from from many different institutions in the United States. Unsurprisingly, almost all of the associate professors, graduate students, and postdocs were at Ph.D.-granting institutions (Fig.~\ref{fig:insttypes}). The undergraduate students came from more diverse institution types, with about half at Ph.D.-granting institutions and the other half at primarily undergraduate institutions.

\subsection{Data analysis}

We first created stacked bar charts of participant responses to the Likert item to gain a baseline understanding of how each population perceived the impacts of federal changes in science funding to their research (Research Question 1). We examined these response distributions qualitatively, identifying how the majority of respondents answered and noting any substantial differences between populations.

We then conducted a thematic coding analysis of responses to the open-ended item to better understand the nature of these perceptions and identify the specific impacts to research (Research Question 2). The first two authors individually read through all of the responses to get a sense of the data~\cite{Tesch1990}. We then met to discuss common themes and ideas and drafted an inductive codebook. That is, the codes were emergent from the data rather than based on prior literature or a previously developed codebook. At this stage, we also grouped together similar codes into two overarching themes: financial impacts and personnel impacts~\cite{saldana2021coding}. The first two authors then iteratively modified the codebook by individually coding a subset of the data (30 responses, 10 from each survey population, because we expected professors, graduate students, postdocs, and undergraduate students to describe these impacts in different ways) and then meeting to improve the code
definitions based on coding disagreements~\cite{Campbell2013}. We determined intercoder reliability by calculating Krippendorff's alpha between the two coders because each response could receive multiple
codes~\cite{krippendorff2011computing}. We reached a Krippendorff's alpha of 0.83, which indicates sufficient reliability because this value is above the commonly used threshold of 0.8~\cite{watts2021current}. While this agreement provides evidence that our codebook is robust to different coders, having one author code all of the responses after reaching sufficient intercoder reliability limits the perspectives represented in the final coding. Therefore, we coded to consensus by having the first two authors each individually code all responses and then meet to negotiate discrepancies until full agreement was reached~\cite{pfeifer2023venturing}. 
With the consensus codes, we qualitatively examined code frequencies across the full dataset (i.e., the fraction of all responses containing each code) and across the three populations to identify the most commonly mentioned impacts in total and for each career stage. 

\section{Results}

We present the results by research question. First, we summarize responses to the Likert item. Then, we describe the impacts that emerged from our inductive, thematic analysis using example responses from each survey population and examine the frequencies of the mentioned impacts across populations.

\subsection{Most respondents view the impacts as negative}

\begin{figure}[b]
    \centering
    \includegraphics[width=3.4in, trim = {0 0.75cm 0 0 }]{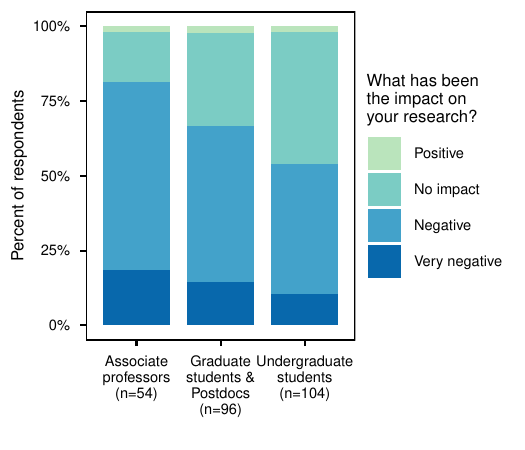}
    \caption{Percent of respondents from each population (Table~\ref{info}) that selected each response to the Likert survey item. No respondents selected ``very positive." 
    }
   
    \label{fig:main}
\end{figure}

More than half of the respondents within each survey population (i.e., associate professors, graduate students and postdocs, and undergraduate students) reported that the impacts of federal changes to science funding on their physics research have been ``negative" or ``very negative" (Fig.~\ref{fig:main}). One associate professor, two graduate students/postdocs, and two undergraduate students responded ``positive" to the Likert item. Three of these five respondents indicated negative impacts in their response to the open-ended item, suggesting that these ``positive" responses to the Likert item may have been due to misinterpretation or mistake. There also appears to be an association between responses to the Likert item and career stage, with associate professors perceiving the impacts most negatively, followed by graduate students/postdocs and then undergraduate students (Fig.~\ref{fig:main}). 

\begin{table*}[t]
\caption{Codebook used to characterize open-ended responses. Example responses for each code are provided in the main text (see Sec.~\ref{exresponses}). \label{codes}}
\setlength{\tabcolsep}{5
pt}
\renewcommand{\arraystretch}{1.3}
\begin{tabular}{ll}
\hline
\hline
Category or code &
Definition  \\
\hline
\textbf{Financial impacts} \\
\hspace{2mm} \parbox[t]{4cm}{\raggedright Current funding}
& \parbox[t]{13cm}{\raggedright Respondent describes losing existing sources of physics research funding, having existing sources of research funding suspended, being declined funding for projects that likely would have been funded under other circumstances, and/or running out of funding.}
\\
\hspace{2mm} \parbox[t]{4cm}{\raggedright Future funding}
& \parbox[t]{13cm}{\raggedright Respondent describes concerns about the ability to secure research funding in the future, including renewals of current grants, and/or to sustain their current level of research productivity in the future.
}
\\
\hspace{2mm} \parbox[t]{4cm}{\raggedright Operational costs}
& \parbox[t]{13cm}{\raggedright Respondent describes heightened costs (e.g., due to tariffs) and/or inability to cover costs of conducting research, such as supplies and equipment, relevant administrative staff, and/or shared spaces within their institution (e.g., cleanroom).
}
\\
\textbf{Personnel impacts}\\
\hspace{2mm} \parbox[t]{4cm}{\raggedright Job security}
& \parbox[t]{13cm}{\raggedright Respondent describes issues with recruiting, hiring, or retaining research group members and/or concerns about future job or education opportunities in physics (e.g., professor roles, graduate programs).
}
\\
\hspace{2mm} \parbox[t]{4cm}{\raggedright New roles}
& \parbox[t]{13cm}{\raggedright Respondent describes having to find different or additional sources of funding, such as through teaching assistantships.}
\\
\hspace{2mm} \parbox[t]{4cm}{\raggedright Mental health}
& \parbox[t]{13cm}{\raggedright Respondent describes experiencing negative emotions or negative impacts to mental health or wellness (e.g., anxiety, stress, and lack of motivation).
}
\\
\hspace{2mm} \parbox[t]{4cm}{\raggedright International researchers}
& \parbox[t]{13cm}{\raggedright Respondent describes concerns about international researchers, such as those related to travel, hiring, and visas.
}
\\
\hspace{2mm} \parbox[t]{4cm}{\raggedright Professional development}
& \parbox[t]{13cm}{\raggedright Respondent describes fewer opportunities to engage in professional development activities, such as presenting at research conferences, contributing to outreach programs, participating in extracurricular activities, and mentoring.
}
\\
\hline
\end{tabular}
\end{table*}

\subsection{Respondents identify financial and personnel impacts}
\label{exresponses}

On the open-ended item, respondents discussed both impacts to research funding (\textit{financial impacts}) and impacts to the people conducting research (\textit{personnel impacts}, Table~\ref{codes}). In the next two subsections, we provide example responses related to these impacts from each survey population to highlight the voices of physics researchers from various career stages. In the third subsection, we present our descriptive analysis of the fraction of respondents that mentioned each of these impacts. 

\subsubsection{Financial impacts}

\textbf{Current funding.} Respondents described losses of current research funding, often in the form of grant cancellations. One associate professor wrote, \textit{``I have had two awards suspended at the end of March 2025. There is no indication these will return, and those each go through 2026, meaning up to 18 months of lost funding for each award."} Graduate students, postdocs, and undergraduate students similarly described that their research advisor(s) lost existing grants. One undergraduate student reported, \textit{``the lab(s) I have been working in have either had complete cuts to funding (Department of Defense) for certain projects or have received less funding than in previous years."} Some graduate students/postdocs also noted that they lost their own fellowship funding, for instance: \textit{``I lost my previous fellowship."}

\textbf{Future funding.} Respondents felt concerned about securing research funding in the future. Several associate professors discussed grant reviews being delayed, grant renewals being declined, and upcoming funding opportunities being entirely canceled. For example, one professor mentioned, \textit{``my several funding applications have been pending for a long time due to the federal executive orders."} Another associate professor wrote about the downstream effects of such delays and cancellations: \textit{``I am worried about sustaining momentum in my research program."} Graduate students and postdocs acknowledged that their research advisor's grants impact their own funding situation, as graduate research assistantships and postdoc positions are often fully supported by federal grants. One trainee said, \textit{``I am uncertain as to how the next few years will look and if my funding is guaranteed in the future."} Graduate students, postdocs, and undergraduate students also discussed that some funded research opportunities they hoped to participate in are no longer available due to lack of funding. One undergraduate student responded, \textit{``REU (Research Experiences for Undergraduates) opportunities I have applied to have been canceled."}

\textbf{Operational costs.} Across career stages, respondents described that administrative changes increased the operational costs of conducting physics research and/or hindered their ability to cover these costs. One associate professor mentioned \textit{``additional tariff charges to our equipment orders."} A graduate student/postdoc described that being unable to cover research costs decreases research productivity: \textit{``one of the biggest challenges is that when instruments in the lab break, there isn't always funding available to repair or replace them quickly. This means experiments can be delayed for long periods of time, which slows down progress."}  Similarly, an undergraduate student noted, \textit{``during the summer, the research [team] couldn't access all the materials needed because of the loss of funding, which limited the experiments we were able to perform."}

\begin{figure*}[t]
    \centering
    \includegraphics[width=7in, trim = {0 0.65cm 0 0}]{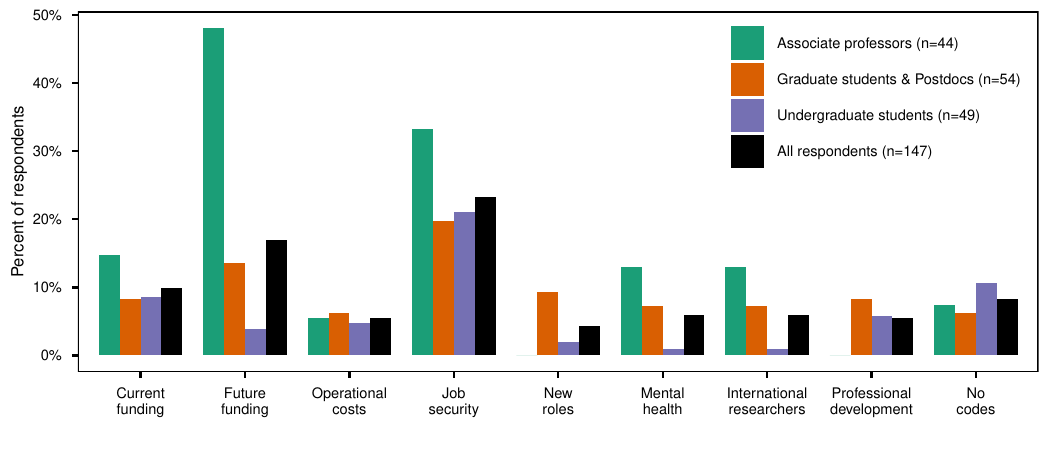}
    \caption{Percent of respondents from each survey population, and across all respondents, whose open-ended response contained each code. Multiple codes could be applied to each response. ``No codes" indicates that zero codes were applied to the response.
    }
   
    \label{fig:codesagg}
\end{figure*}

\subsubsection{Personnel impacts}

\textbf{Job security.} Many respondents discussed concerns related to employment. Associate professors were primarily concerned about hiring researchers for their lab and the future job prospects of their research group members, rather than their own job stability. For some associate professors, these concerns manifested in changes to their future hiring plans: \textit{``the uncertainty in funding makes me hesitant to recruit new team members."} In other cases, professors were already downsizing: \textit{``I have decided to start consolidating and reducing my team, as I am uncertain whether I will be able to fund future students through the full PhD program. So, for example, I have decided not to take new students/postdocs this year."} Graduate students, postdocs, and undergraduate students, by contrast, expressed significant uncertainty about their own future job opportunities. One graduate student wrote, \textit{``my contract ended this summer, and the uncertainty in funding has made it difficult to plan the next steps in my career. While I am still committed to continuing my research, the shifting funding landscape has made me question whether it is worthwhile to apply for postdoctoral fellowships at this time, given the instability and reduced opportunities."} Undergraduates were primarily worried about graduate program admissions, for example: \textit{``there have been significantly less opportunities for pursuing graduate programs."}

\textbf{New roles.} Relatedly, graduate students, postdocs, and undergraduate students described having to pivot roles or research areas and/or take on supplementary jobs. For graduate students, role changes often took the form of picking up teaching assistantships when research assistantships were no longer possible due to federal funding cuts. One student stated, \textit{``rather than being able to not teach anymore and focus solely on research, I'm more or less forced to now teach to make up for funding gaps."} An undergraduate student discussed that they were no longer able to be compensated for their research duties, so they took on additional roles to make up for it: \textit{``my ability to receive pay [has] been majorly hindered and thus I have had to utilize outside jobs to be able to support myself through research."}

\textbf{Mental health.} Respondents also mentioned substantial negative impacts to physics researchers' mental health and well-being, such as stress and declining morale. One associate professor noted, \textit{``the general existential dread and uncertainty is extraordinarily taxing. And the foreign students range from annoyed to demoralized to, in one case, experiencing severe mental health impacts."} Similarly, a graduate student/postdoc described, \textit{``these changes create uncertainty in achieving my future goals. That [has] shaken my motivation and focus, thinking of lacking opportunities in future."} Another graduate student/postdoc described impacts to their entire research group's culture: \textit{``it affected the group atmosphere, making everyone paranoid and angry--it was my worst summer since I started studying."} One undergraduate student mentioned the noticeable emotional toll on faculty research advisors: \textit{``we can sense our professors are afraid."}

\textbf{International researchers.} Some respondents specifically noted hardships for international researchers, such as travel restrictions and strict hiring policies. These responses often overlapped with those about mental health. An associate professor responded, \textit{``my lab is a very international team, and the statements and actions from the administration have significantly dampened their morale. It's hard to be motivated to do your best work when the president says he is planning to `aggressively revoke' your visa."} One graduate student wrote, \textit{``these recent changes have definitely [affected] my mental peace about my PhD journey, I feel an urgency now to finish as much research as I can before there is some issue with visa laws or an effect on my funding situation."} Related to job opportunities for international researchers, one undergraduate respondent said, \textit{``due to funding issues, the Fermilab SIST (Summer Internships in Science and Technology) internship is not running anymore and as an international student I am only eligible for SIST internship and not the SULI (Science Undergraduate Laboratory Internship) internship."}

\textbf{Professional development.} Graduate students, postdocs, and undergraduate students highlighted new limitations to professional development opportunities, such as attending conferences and receiving mentorship from their research advisors. One graduate student/postdoc noted that \textit{``limited funding makes it difficult to attend conferences, which are vital for presenting my work and building collaborations."} An undergraduate student mentioned that \textit{``faculty cuts have impacted my access to knowledge/mentorship."} Other undergraduate students described new obstacles to participating in service or extracurricular opportunities, for example: \textit{``for outreach [organizations] that I am a part of, it is basically impossible to get funding to support events."}

   

\subsubsection{Descriptive analysis}

With the exception of finding new roles and reduced professional development opportunities, all of the above impacts were described by at least one respondent from all three career stages in our dataset (Fig.~\ref{fig:codesagg}). Across all respondents, the most frequently mentioned impacts were job insecurity, future funding concerns, and loss of current funding (black bars in Fig.~\ref{fig:codesagg}). In addition to these three impacts, negative impacts to mental health and concerns about international researchers were salient for the associate professors (green bars in Fig.~\ref{fig:codesagg}). Graduate students and postdocs primarily discussed these three prevalent impacts as well, but they also focused on new roles and professional development (orange bars in Fig.~\ref{fig:codesagg}). Undergraduate students mainly discussed job security, losses of current funding, and professional development, with fewer undergraduates mentioning concerns about future research funding than those in more advanced career stages (purple bars in Fig.~\ref{fig:codesagg}). 

The degree to which respondents described these impacts also varied by career stage. For those who responded to the open-ended survey item, associate professors, graduate students and postdocs, and undergraduate students wrote an average of 49, 39, and 28 words, respectively, and discussed an average of 1.6, 1.4, and 1.0 codes, respectively. 

These descriptive patterns provide further insight to the largely negative responses to the Likert item (Fig.~\ref{fig:main}) and indicate variation in impacts across different academic populations. Notably, there are no substantial differences in mentioned impacts when the results are also disaggregated by Likert item response (Fig.~\ref{fig:codesdisagg} in the Appendix).

\section{Discussion}

In this study, we administered surveys to physics undergraduate students, graduate students, postdocs, and associate professors asking them about the impacts of recent changes to the federal science funding landscape on their research. We found that most respondents viewed the impacts as negative. Respondents documented various financial and personnel impacts, including loss of current funding, uncertainty about future funding, job insecurity, negative effects to mental health, and concerns about international researchers. These results corroborate those of previous studies aiming to understand the effects of recent federal funding changes to scientific research~\cite{white2025impacts,bennett2025advocacy,Hammond2025trump} and suggest that these existing findings hold across a range of career stages, from undergraduate students to associate professors.

We also identified several differences in the survey responses across career stages. On the Likert item, the most negative perceptions came from associate professors, followed by graduate students and postdocs and then undergraduate students. There are several possible explanations for this pattern. First, it could be due to the timing of each survey (Table~\ref{info}). Associate professors responded the earliest (July 2025), closest to when these federal changes, particularly the executive orders, were first announced. Graduate students and postdocs responded later (September 2025), and undergraduate students responded last (January 2026), after the new budget for the NSF had been approved with relatively minimal cuts. As a result, earlier respondents (i.e., the associate professors) may have been reacting to a period of greater uncertainty about the scope of potential funding reductions.

Alternatively, or in addition, this pattern may reflect differences in awareness of the scientific funding landscape across career stages. Associate professors frequently apply for and manage grants from federal agencies and therefore must closely monitor funding priorities and policy developments. Graduate students and postdocs are often supported by these grants and may discuss them with their advisors, but they typically have less direct responsibility for proposal preparation and grant management. Undergraduate students, by contrast, are more commonly supported through internal university funds or dedicated undergraduate research programs (e.g., NSF Research Experiences for Undergraduates) and/or may have fewer conversations with faculty research advisors about the details of current grant funding. Consequently, undergraduates may be less directly exposed to changes in federal funding policy and their potential implications, which may explain the relatively large fraction of undergraduates who responded to the Likert item with ``no impact."

On the open-ended item, associate professors were primarily concerned about existing and future funding opportunities, while many graduate students, postdocs, and undergraduate students were concerned about future job prospects. Similar to the Likert item, these differences may be attributable to survey timing (i.e., associate professors filled out the survey the earliest and closest in time to the first executive orders) and/or career stage itself (i.e., associate professors have tenure and are less likely to be looking for a new job). As trainees at lower career stages are more distant from the grant funding and management processes, these populations may not be familiar enough with these processes in general and with their particular research group's funding situation to know about and articulate specific impacts to research. Indeed, we observed that early career researchers wrote fewer words and mentioned fewer impacts in their responses than more senior researchers. Future studies should continue to examine the relationship between career stage and reported impacts to research, for example through additional survey prompts and/or interviews.

We also noticed that many of the impacts measured on the open-ended item do not occur in isolation. At a high level, many respondents mentioned at least one impact on the open-ended item, indicating that individual researchers are experiencing a range of effects. When considering the impacts themselves, we start to uncover an interwoven web of negative impacts as a result of the changing funding landscape. In some cases, combinations of impacts form recursive cycles. For example, several graduate students mentioned that they lost their research assistantship and needed to acquire funding through a teaching assistantship; meanwhile, associate professors acknowledged that institutional spending cuts have led to fewer teaching assistantships being available in their departments. As another example, a few undergraduate students reported a decrease in access to their mentors; meanwhile, some associate professors mentioned increased difficulty in mentoring their research students. Future research should further investigate the interplay between different impacts, for example by using mixture models~\cite{muthen2001latent,masyn2013latent-key} to identify common combinations of impacts reported on surveys or by leveraging revealed causal mapping~\cite{nelson2000revealed,scanlon2019method} with interview data to identify causes and effects.

Another finding worth highlighting is that concerns about international researchers represent a somewhat distinct category of impact compared to the other effects. While most of the impacts we documented stem directly or indirectly from reductions in federal research funding, worries about international scholars arise primarily from changes in immigration and visa policy~\cite{mervis2025visa}. This distinction between funding-related and policy-related impacts matters because each type may require different advocacy responses and the effects on international researchers may persist or intensify even when federal research funding is restored. Future work should consider treating funding versus policy impacts as a separate dimension of analysis.

One final takeaway from our data is that respondents were concerned about funding opportunities and personnel more than outright research productivity. Only a few respondents indicated concern about their research projects being delayed, and no respondents mentioned a decrease in the number of deliverables (e.g., publications and presentations) they were able to produce. Associate professors in particular expressed deep worry about the job prospects and well-being of their advisees. This pattern reflects that scientific research is only as strong as the human infrastructure supporting it. Faculty concerns for their trainees points to a threat that is not only operational, but also generational~\cite{langin2025earlycareer,langin2026pressure}. We hope that future work will examine the longitudinal effects of these administrative changes to personnel, especially early career physics researchers. 


\subsection{Potential implications}

Our results should be considered exploratory given the limited existing research about such recent events.  However, our study points to several possible implications to support the development of the research training pipeline in physics. As noted above, one opportunity is related to faculty mentoring. These federal funding changes are not only new to research trainees, but also to faculty members who are re-navigating their research and mentoring responsibilities. As undergraduate and graduate students' relationships with their supervisors comprise a key factor in their retention~\cite{limeri2019mentor,Sim2026}, researchers should continue to probe the kinds of mentoring resources that would be most useful to faculty members, and relevant organizations and agencies should consider how to coordinate mentor training opportunities. Based on the current study, such training may provide faculty members with updated resources related to student mental health challenges and career paths. The Effective Practices for Physics Programs guide may serve a central role in disseminating these resources~\cite{ep3guide,craig2025helping}.

Graduate student and postdoc data suggest that these groups need the most support in identifying funding sources (i.e., obtaining research versus teaching assistantships and deciding between research groups) and navigating their career trajectories. Previous research has systematically documented the process through which graduate students join a research group and devised suggestions for physics departments to facilitate this process ~\cite{verostek2024modeling,verostek2024physics}. Our data indicate that these recommendations may need to be modified in light of the shifting funding landscape, with a larger focus on transparency and stability of funding. Physics departments and individual faculty advisors may also need to become more equipped to prepare their graduate and postdoc advisees for non-academic careers, as the faculty job market is shrinking~\cite{langin2025earlycareer}. 

Undergraduate students reported facing a significant decrease in research opportunities (e.g., REU cancellations), precluding them from gaining the skills needed to pursue graduate school or a non-academic, research-oriented career~\cite{russell2007benefits}. One opportunity for undergraduate physics instruction, therefore, is to re-consider course structures such that students develop research skills in their courses instead of or in addition to faculty members' research groups. It has been shown, for example, that discovery-oriented physics labs and course-based undergraduate research experiences improve students' critical thinking skills more than traditional lab courses~\cite{smith2023course,holmes2015teaching,walsh2022experimentation,werth2022impacts}. Moreover, federal changes have only tightened scientific funding, which may exacerbate inequities in the sociodemographic backgrounds of students who have access to undergraduate research experiences through faculty members' grants. Research-based labs and courses have been shown to mitigate these inequities~\cite{bangera2014course}.

\subsection{Limitations}

There are several important limitations of this study that constrain our scope of inference and prompt future research studies. First, we collected survey data at different timepoints for each population, limiting the direct comparability of responses across career stages. As the national funding landscape continues to evolve, more controlled studies measuring the impacts to different populations at similar timepoints will validate and extend our findings. Such studies will also illuminate whether and how the impacts identified here dissolve or persist as funding is reinstated and individuals try to rebuild disrupted projects and career plans.

We did not collect information about respondents' research areas, though it is plausible that differences in impacts exist across physics subfields. For example, grants related to climate, education, and diversity, equity, and inclusion have been disproportionately at risk for getting terminated~\cite{gewin2025stemdivefforts,crockett2025nihnsf}. We encourage future research to disaggregate impacts across subfields.

Our data include responses from associate professors, but not assistant or full professors. However, assistant professors are likely experiencing more and/or different concerns related to funding than associate and full professors, for example to fulfill tenure and promotion requirements. Additionally, all associate professors in this study were experimental physicists, limiting generalizability to associate professors conducting theoretical research. All associate professors studied here also received a large funding award from the Gordon and Betty Moore Foundation, a philanthropy that is far less susceptible to administrative changes than federal agencies such as the NSF and the Department of Energy. Therefore, the impacts reported by faculty members in this analysis may differ from the impacts experienced by physics faculty members broadly. On one hand, the associate professors in this study may experience fewer impacts than the broader population of associate professors because they have significant funding from a private philanthropy. Alternatively, the associate professors in this study may face more impacts than the broader population of associate professors if they also hold several federally funded awards, given their prolific research profiles. We recommend that researchers collect similar data from physics professors at all levels and across different types of research that may rely on different organizations for research funding.

Finally, responses for the undergraduate students, graduate students, and postdocs were collected via registration surveys for conferences aimed at women and gender minorities in physics. We recognize that this limits the demographic backgrounds and, therefore, perspectives that are represented in our data for these career stages; however, we also acknowledge the strength in voicing these typically marginalized viewpoints in physics~\cite{clark2005,sax2016}. Further work should consider survey data collection methods that reach a wider range of student and postdoc backgrounds.

\section{Conclusion}

This study only begins to shed light on the range of impacts of federal changes in science funding to academic physics researchers. We identify predominantly negative impacts experienced by undergraduate students, graduate students, postdocs, and associate professors, including concerns about funding opportunities, job security, mental health, and international scholars. We plan to continue tracking these effects to researchers at various career stages as the funding landscape continues to evolve. We hope that further research builds on our analysis in order to identify the best forms of support for research advisors and trainees, both in physics and in other scientific disciplines.

\section*{ACKNOWLEDGEMENTS}

We thank David Esparza and Ibukunoluwa Bukola for their feedback on this manuscript. This material is based upon work supported by the Gordon and Betty Moore Foundation and the American Physical Society.

\bibliography{Funding} 


\section*{APPENDIX}

\subsection{Descriptive analysis disaggregated by Likert response}

\begin{figure*}[t]
    \centering
    \includegraphics[width=7in, trim = {0 0.5cm 0 0}]{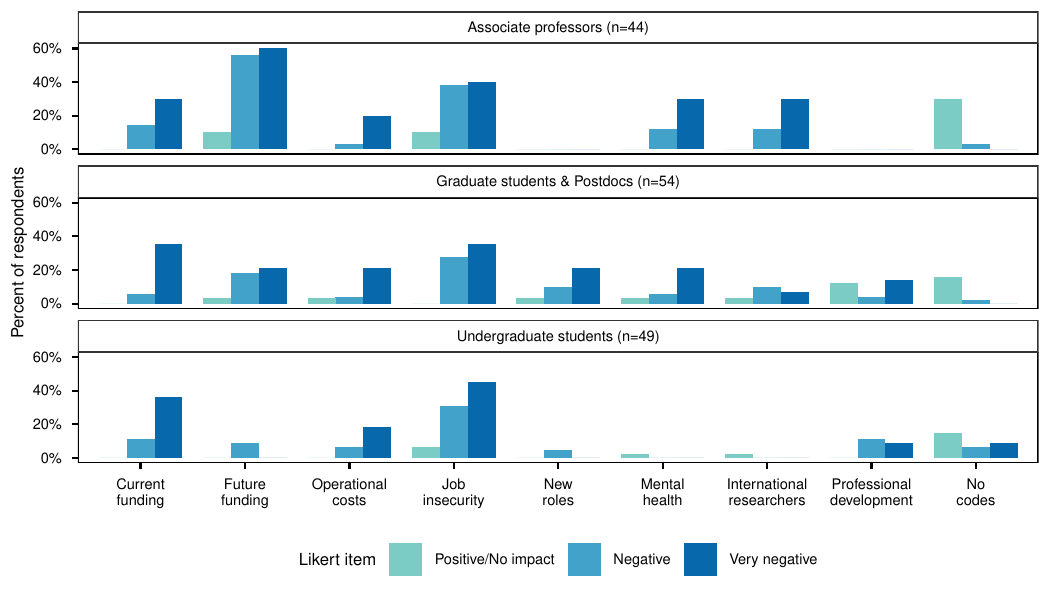}
    \caption{Percent of respondents from each survey population, and for each Likert item response, whose open-ended response contained each code. We collapsed ``positive" and ``no impact" because a very small fraction of respondents selected ``positive" on the Likert item (Fig.~\ref{fig:main}). Multiple codes could be applied to each response. ``No codes" indicates that zero codes were applied to the response.
    }
    \label{fig:codesdisagg}
\end{figure*}

Most respondents who indicated ``positive" or ``no impact" on the Likert item did not mention any impacts captured by our codebook on the open-ended survey item (Fig.~\ref{fig:codesdisagg}). Only a few differences between those who responded with ``negative" versus ``very negative" emerge, and these findings should be considered preliminary given that respondents may interpret Likert scales differently. For example, especially among graduate students and postdocs, respondents who indicated ``very negative" on the Likert item mentioned impacts to current funding more frequently than those who indicated ``negative." Respondents who indicated ``very negative" on the Likert item also mentioned operational costs and mental health more frequently than those who indicated ``negative." Among associate professors, concerns about international researchers were more salient for those who responded ``very negative" to the Likert item than the other Likert item response options. These observations are qualitative and require further research to make quantifiable claims.

\end{document}